\documentclass[aps,preprint,floatfix,showpacs,showkeys]{revtex4}
\usepackage{epsf,amsmath,amssymb,verbatim,color,bm}
\begin{document}
\newcommand{\be}{\begin{equation}}
\newcommand{\ee}{\end{equation}}

\title{Self-similarity in Fractal and Non-fractal Networks}
\author{J.~S. Kim}
\author{B. Kahng}
\email{kahng@phya.snu.ac.kr}
\author{D. Kim}
\affiliation{Center for Theoretical Physics {\rm \&} Frontier Physics Research Division, 
Department of Physics and Astronomy, Seoul National University, Seoul 151-747, Korea}
\author{K.-I. Goh}
\affiliation{Department of Physics, Korea University, Seoul 136-713, Korea}
\date{\today}

\begin{abstract}
We study the origin of scale invariance (SI) of the degree
distribution in scale-free (SF) networks with a degree exponent
$\gamma$ under coarse graining.
A varying number of vertices belonging to a community or a box in
a fractal analysis is grouped into a supernode, where the box mass $M$
follows a power-law distribution, $P_m(M)\sim M^{-\eta}$. The
renormalized degree $k^{\prime}$ of a supernode scales with its box
mass $M$ as $k^{\prime} \sim M^{\theta}$. The two exponents $\eta$
and $\theta$ can be nontrivial as $\eta \ne \gamma$ and $\theta <
1$. They act as relevant parameters in determining the
self-similarity, i.e., the SI of the degree distribution, as follows:
The self-similarity appears either when $\gamma \le \eta$ or under
the condition $\theta=(\eta-1)/(\gamma-1)$ when $\gamma> \eta$,
irrespective of whether the original SF network is fractal or
non-fractal. Thus, fractality and self-similarity are disparate
notions in SF networks.
\end{abstract}

\maketitle

\section{Introduction}

Kadanoff's block spin and coarse-graining (CG) picture is the
cornerstone of the renormalization-group (RG)
theory~\cite{kadanoff}. A system is divided into blocks of equal
size and is described in terms of the block variables that
represent the average behavior of each block. Scale invariance at
the critical point under this CG enables one to evaluate the
critical exponents. From a geometric point of view, scale
invariance implies the presence of fractal structures, and their
fractal dimensions are associated with the critical
exponents~\cite{nijs}. For example, the magnetization and the
singular part of the internal energy for the critical Ising model
are such scale invariant quantities, which are manifested as the
fractal geometric forms of the area of spin domains and the length
of the spin domain interface, respectively.

The fractal dimension $d_B$ of a fractal object is measured by
using the box-covering method~\cite{feder}, in which the number of
boxes, $N_B(\ell_B)$, needed to tile the object with boxes of size
$\ell_B$ follows a power law,
\begin{equation}
N_B(\ell_B)\sim \ell_B^{-d_B}. \label{fractal}
\end{equation}
This relation is referred to as {\em fractal scaling} hereafter. A
fractal object is self-similar in the sense that it contains
smaller parts, each of which is similar to the entire object~\cite{feder}.

\begin{figure*}
\centerline{\epsfxsize=15.0cm \epsfbox{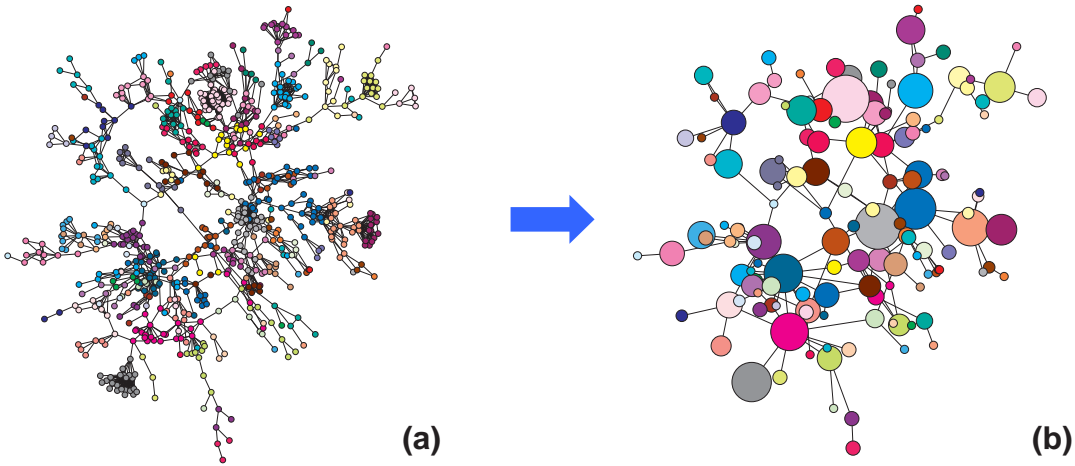}} 
\caption{ (Color online) (a) The protein interaction network of the budding yeast and
(b) its coarse-graining. Group masses $\{M_{\alpha}\}$, detected by
the box covering, are heterogeneous. The node size in (b) is taken as
$\sim \sqrt{M_{\alpha}}$ for visualization.} \label{protein}
\end{figure*}

While the notions of CG and self-similarity are well established
when an object is embedded in Euclidean space, it is not clear how
to extend and apply them usefully to objects not embedded in
Euclidean space. In the former case, the number of vertices within
each box, equivalent to the block size in RG terminology and
referred to as the group mass for future discussion, is almost
uniform, and the fractal objects are self-similar and vice versa. In
the latter case, however, the group mass is extremely heterogeneous
so that it follows a power-law distribution. This case can happen in
scale-free (SF) networks; then, one needs to establish the way of
CG and the notion of self-similarity in a new setting, which is a
goal of this paper. A SF network~\cite{ba} is a network whose
distribution $P_d(k)$ of degree $k$, the number of edges connected
to a given vertex, follows a power-law form: $P_d(k)\sim
k^{-\gamma}$ with the degree exponent $\gamma$. The scale invariance
of the degree distribution under CG is defined as the
self-similarity in SF networks.

Most SF networks in the real world contain functional groups or
communities within them~\cite{hmodel}. In general, the distribution
of the group mass $M$, the number of vertices within each group follows
a power law asymptotically~\cite{newman,albert}:
\begin{equation}
P_m(M)\sim M^{-\eta}. \label{gsize}
\end{equation}
When such groups are formed within networks, it would be more
natural to take each group as a unit to form a supernode in CG
because the vertices within each group are rather homogeneous in their
characteristics, such as functionality or working division, and are
connected densely. In other CG procedures, groups may not
necessarily represent functional modules in bio-networks and
communities in social networks; they can be taken arbitrarily in a
theoretical perspective, for example, being artificially composed or
boxes introduced in the fractal
analysis~\cite{ss,jskim,chaos,newjournal}. Supernodes are connected
if any of their merging vertices in different communities are
connected. This CG method is different from the standard ones in
view of extremely heterogeneous group masses as shown in
Fig.~\ref{protein}. We show below that the exponent $\eta$ plays a
central role in determining the self-similarity.

So far, several endeavors to achieve a RG transformation for SF
networks have been carried out. However, their methods remain in the
framework of the standard RG method, ignoring the heterogeneity of
the group-mass distribution. For example,
Kim~\cite{bjkim,bjkim_jkps} applied CG to a SF network generated on a
Euclidean space~\cite{rosenfeld}, including long-range edges. Taking
advantage of the underlying Euclidean geometry, the number of
vertices within each block is uniform and increases in a power law
as the block lateral size increases, so that the real-space RG
method can be naturally applied. In Refs.~\cite{mendes} and \cite{jung}, the
decimation method was applied to a few deterministic SF networks,
which were constructed recursively starting from each basic
structure. Those deterministic models restore their shapes under CG
achieved by decimating the vertices with the smallest degree at each
stage. We discuss this case in detail later.

In this paper, we perform the CG of SF networks in three different
ways: (i) random grouping, (ii) box covering in fractal analysis,
and (iii) identifying community structure with clustering
algorithms. The groups are (i) artificially composed, (ii) taken as boxes
introduced in the fractal analysis, and (iii) taken as communities
embedded within the network, in respective cases. For all cases, the group-mass
distribution follows the power law, Eq.~(\ref{gsize}). We identify the
criteria for self-similarity in terms of the exponents $\eta$ in Eq.~(\ref{gsize})
and $\theta$ which will be introduced below in Eq.~(\ref{nonlinear}). 
Such a self-similarity condition holds for non-fractal, 
as well as fractal real-world networks.

\section{CG by random grouping}

\begin{figure}
\centerline{\epsfxsize=7.5cm \epsfbox{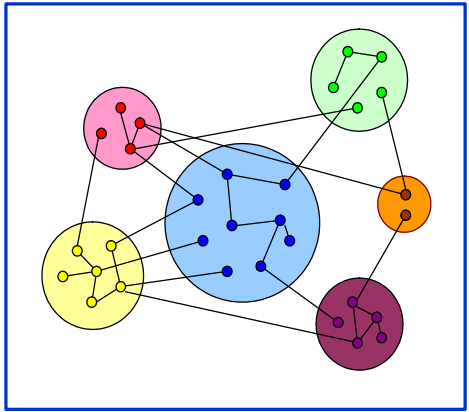}} 
\caption{ (Color online) Schematic snapshot of a random SF network with random
grouping.} \label{group}
\end{figure}

\begin{figure}
\centerline{\epsfxsize=8.5cm \epsfbox{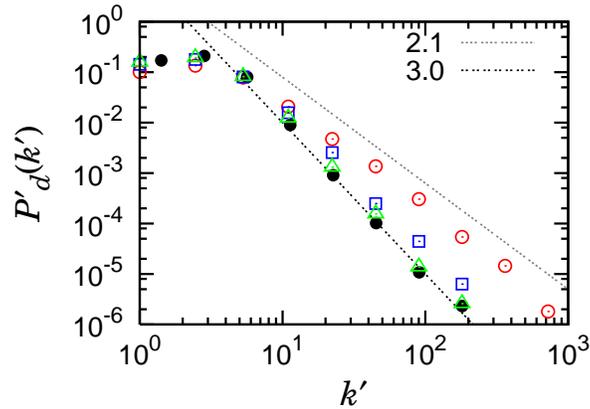}} 
\caption{ (Color online) Degree distributions of a renormalized network of the static
model for $\gamma=3$ under CG. The group masses are preassigned to
follow a power law $P_m(M)\sim M^{-\eta}$ with
$\eta=2$(\textcolor{red}{$\circ$}), 3 (\textcolor{blue}{$\square$}),
and 4(\textcolor{green}{$\triangle$}). The degree distribution of
the original network is also drawn ($\bullet$). The dashed and the dotted
lines with slopes of 2.1 and 3.0, respectively, are drawn as
guidelines.} \label{static}
\end{figure}

The model enables us to obtain the renormalized degree exponent
$\gamma^{\prime}$ analytically by using the generating function
technique, showing that, indeed, $\gamma^{\prime}$ depends on
$\eta$. The model is constructed as follows: (i) We construct a SF network
through the static model of Ref.~\cite{static_model}. We choose the
degree exponent $\gamma=3$ and the mean degree $\langle k \rangle
=4$. (ii) $N$ individual vertices are grouped randomly into
$N^{\prime}$ groups with sizes $\{M_{\alpha}\}$
($\alpha=1,\dots,N^{\prime}$), following the distribution function in Eq.~(\ref{gsize}). 
The exponent $\eta$ is tuned. We consider three
cases of $\eta$: (a) $\eta=2~(< \gamma)$, (b) $\eta=3~(=\gamma)$,
and (c) $\eta=4~(> \gamma)$. (iii) We perform the CG by replacing
each group with a supernode and connecting them if any of their
merging vertices in different groups are connected. A renormalized
network is constructed. A schematic snapshot of a constructed network
is shown in Fig.~\ref{group}. Next, the degree distribution
$P^{\prime}_{d}(k^{\prime})\sim k^{\prime -\gamma^{\prime}}$ of the
renormalized network is measured. The result is as follows:
$\gamma^{\prime}=\eta=2\ne \gamma$ for $\eta=2$ (a),
$\gamma^{\prime}=\gamma=3$ for $\eta=3$ (b), and
$\gamma^{\prime}=\gamma=3$ for $\eta=4$ (c), as shown in
Fig.~\ref{static}.

The numerical result is understood analytically as follows: Since
the vertices are grouped randomly, every vertex has an equal
probability to connect to vertices in other groups per edge, so the
total probability is proportional to the degree of each vertex. This
leads to the relation $k_{\alpha}^{\prime}\approx \sum_{j\in
\alpha}k_j$, where $\alpha$ is the index of the group. Then, the
degree distribution $P^{\prime}_{d}(k^{\prime})$ of the renormalized
network is written as
\begin{equation}
P^{\prime}_{d}(k^{\prime}) \approx \sum_{M=1}^{\infty} P_m(M)
\sum_{k_1,k_2,\ldots,k_M} \prod_{j=1}^M P_d (k_j) \delta
(\sum_{j=1}^{M} k_j-k^{\prime}),
\end{equation}
$\delta$ denoting the Kronecker delta. By using the generating
function technique, one can find the relation ${\cal
{P}}^{\prime}_{d}(z)={\cal {P}}_m({\cal {P}}_{d}(z))$, where
${\cal {P}}_{d}(z)$ is the generating function of $P_d(k)$ and so
forth. The ${\cal P}_d(z)$ is obtained to be
\begin{equation}
{\cal {P}}_d(z)=1-\langle k \rangle (1-z)+a(1-z)^{\gamma-1}+
\mathcal{O}\left((1-z)^2 \right), \label{generating}
\end{equation}
where $\langle k \rangle=\sum_k k P_d(k)$ and $a$ is a constant. The
generating function ${\cal P}_m(\omega)$ is also derived in a similar
form to Eq.~(\ref{generating}). Then, one can find immediately that\be
\gamma^{\prime}=
\begin{cases}
\gamma, & \text{for}~ \gamma \le \eta,\\
\eta, & \text{for}~\gamma > \eta.
\end{cases}
\label{gammaprimesimple} \ee as long as both $\gamma$ and $\eta > 2$.
Thus, self-similarity holds when $\gamma \le \eta$, and we can
confirm that the exponent $\eta$ plays a key role in determining the
exponent $\gamma^{\prime}$. In the formulation, the relation
$k_{\alpha}^{\prime}\approx \sum_{j\in \alpha}k_j$ was crucial. When
the relation no longer holds, the derivation of the degree exponent
$\gamma^{\prime}$ is more complicated. This can happen for fractal
and some non-fractal networks, and also for clustered networks, which we
discuss next.

\begin{figure}
\centerline{\epsfxsize=10cm \epsfbox{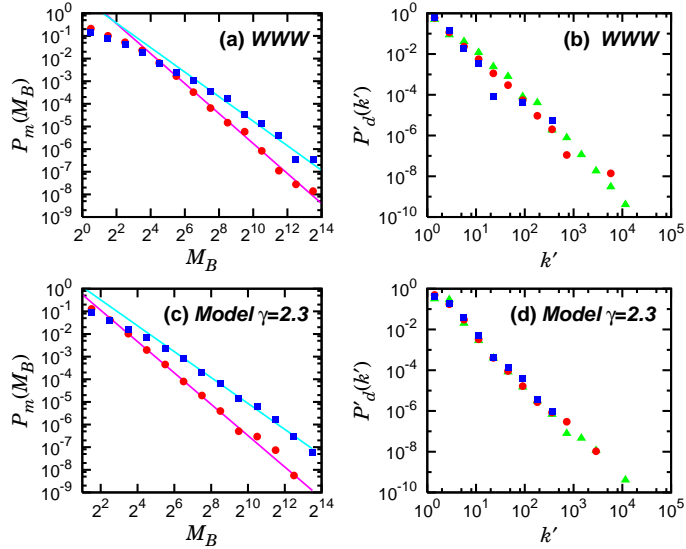}}
\caption{(Color
online) Box-mass distribution for (a) the WWW and (c) the fractal model
network of Ref.~\cite{goh2006}. Data are for $\ell_B=2$
(\textcolor{red}{$\bullet$}) and $\ell_B=5$
(\textcolor{blue}{$\blacksquare$}). The solid lines are guidelines with
slopes of --2.2 and --1.8, respectively, in both (a) and (c). The
degree distributions of the original network
(\textcolor{green}{$\blacktriangle$}) and the renormalized networks
with $\ell_B=2$ (\textcolor{red}{$\bullet$}) and $\ell_B=5$
(\textcolor{blue}{$\blacksquare$}) for (b) the WWW and (d) the fractal
model network. The fractal model has a system size $N\approx
3\times 10^5$. (c) and (d) are adopted from
Ref.~\cite{newjournal}.}\label{www}
\end{figure}

\section{CG by box covering}

Recently, it was discovered~\cite{ss} that fractal
scaling, Eq.~(\ref{fractal}) holds in some SF networks, such as the
world-wide web (WWW), the metabolic network of {\em Escherichia
coli}, and the protein interaction network of {\em Homo sapiens}.
Groups are formed by covering the networks by boxes that contain
nodes whose mutual distances are less than a given box size. The
group-mass distribution follows the power law of Eq.~(\ref{gsize}), even
though the boxes' lateral sizes are fixed as $\ell_B$ for all boxes.
Thus, the fractal networks are good objects for our study.

We first apply a box-covering algorithm to the WWW. Our
box-covering algorithm is slightly modified from the original one
introduced by Song {\em et al.}~\cite{ss}, and the details of the
algorithm is presented in Ref.~\cite{jskim}. Both
algorithms~\cite{ss,jskim} are identical in spirit and the box
size $\ell_B$ we use is related linearly to the corresponding one
$\ell_S$ in Ref.~\cite{ss}. Next, each box is collapsed into a
supernode. Two supernodes are connected if any of their merging
vertices in different boxes are connected. The degree distribution
$P^{\prime}_{d}(k^{\prime})$ of the renormalized network is
examined.

The distribution of box masses is measured and found to follow a
power law asymptotically, $P_m(M)\sim M^{-\eta}$ [Fig.~\ref{www}
(a)]. The exponent $\eta$ is found to depend on the box size
$\ell_B$. For small $\ell_B=1$ or 2, $\eta\approx 2.2$ is measured,
which is close to $\gamma$. On the other hand, as $\ell_B$
increases, we expect $\eta$ to approach the exponent
$\tau=\gamma/(\gamma-1)$, describing the power-law behavior of the
cluster-size distribution of the branching tree~\cite{dslee}. We
find that $\eta \approx \tau \approx 1.8$ for $\ell_B=5$. This
result can be understood as follows: The WWW contains a
skeleton~\cite{skeleton}, a spanning tree based on the betweenness
centrality or load, which can represent the original network in the
box-covering~\cite{goh2006}. For small $\ell_B$, the lateral
dimension of the box is not large enough to see the asymptotic
fractal behavior of the spanning tree, so that the number of
vertices in a given box scales similarly to the largest degree
within that box. Thus, $\eta=\gamma$. As $\ell_B$ increases, on the
other hand, the asymptotic fractal behavior of the spanning tree
becomes dominant. Thus, the exponent $\eta$ becomes the same as the
exponent $\tau$ asymptotically, which was observed in the case of
$\ell_B=5$. The case of $\eta \ne \gamma$ was not considered in
Ref.~\cite{ss}. The origin of the self-similarity is nontrivial as
we show below.

\begin{figure}
\centerline{\epsfxsize=6cm \epsfbox{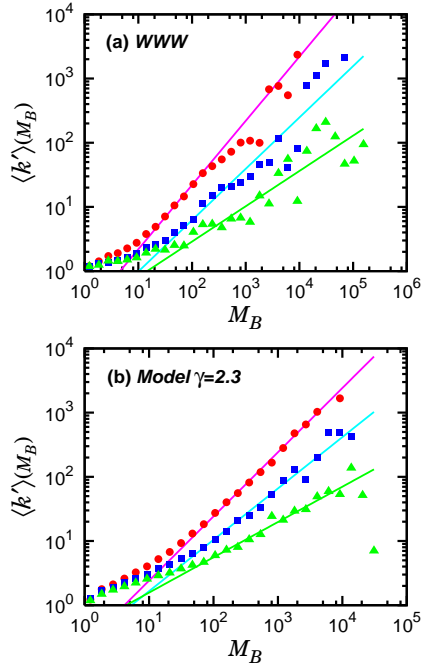}} 
\caption{(Color online) Plot of the average renormalized degree $\langle
k^{\prime}\rangle$ versus box mass $M_B$ for (a) the WWW and (b) the
fractal model network of Ref.~\cite{goh2006}. Data are for box
sizes $\ell_B=2$ (\textcolor{red}{$\bullet$}), $\ell_B=3$
(\textcolor{blue}{$\blacksquare$}), and $\ell_B=5$
(\textcolor{green}{$\blacktriangle$}). The solid lines, guidelines, have
slopes of 1.0 (\textcolor{red}{$\bullet$}), 0.8
(\textcolor{blue}{$\blacksquare$}), and 0.6
(\textcolor{green}{$\blacktriangle$}), respectively, for both (a)
and (b). (b) is adopted from Ref.~\cite{newjournal}.} \label{kprime}
\end{figure}
\begin{figure}
\centerline{\epsfxsize=6cm \epsfbox{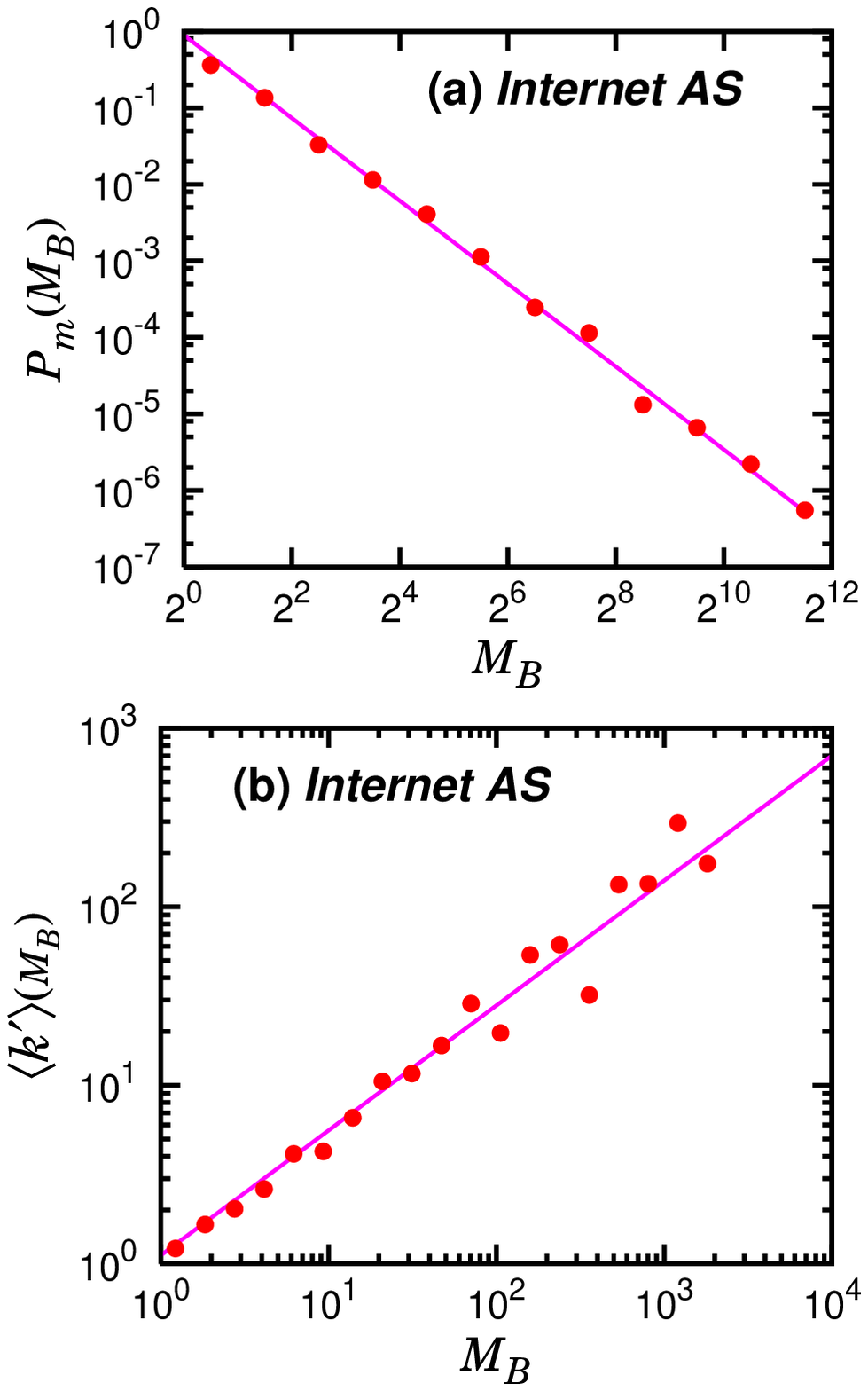}} 
\caption{(Color online) (a) Box-mass distribution and (b) average renormalized degree
$\langle k^{\prime} \rangle$ versus box mass $M_B$ of the
Internet at the AS level. Here, the lateral box size is taken as
$\ell_B=2$ for both. The solid lines, drawn for reference, have slopes
of (a) $-1.8$ and (b) $0.7$.} \label{internet}
\end{figure}

The CG process involves two steps. The first is the
vertex renormalization, i.e., merging of vertices within a box into
a supernode, and the second is the edge renormalization, merging of
multiple edges between a pair of neighboring boxes into a single
edge. The second step can yield a nonlinear relationship between the
renormalized degree $k^{\prime}$ and box mass $M_B$, although the
total number of inter-community edges from a given box is linearly
proportional to its box mass~\cite{jskim}. Thus, we propose that
there exists a power-law relation between the average renormalized
degree and the box mass: \be \langle k^{\prime}\rangle(M_B)\sim
M_B^{\theta}. \label{nonlinear}\ee

The power-law relation Eq.~(\ref{nonlinear}) is tested numerically for
the WWW, as shown in Fig.~\ref{kprime}(a). For $\ell_B=2$, we
estimate $\theta \approx 1.0 \pm 0.1$. Thus, the linear relation
holds. For $\ell_B=3$ and 5, however, $\theta \approx 0.8 \pm 0.1$
and $0.5\pm 0.1$, respectively, implying the nonlinear
relationship, Eq.~(\ref{nonlinear}). One may doubt the nonlinear behavior
due to the scattered data shown in Fig~\ref{kprime}(a). To confirm
this result, we recall a previous study~\cite{newjournal} for the
fractal model with $\gamma=2.3$ introduced by Goh {\em et
al.}~\cite{goh2006}, where we can reduce the data noise by taking
an ensemble average over network configurations. We obtained similar
results as shown in Fig.~\ref{kprime}(b). The data for the fractal
model are averaged over 10 different network configurations, so 
the nonlinear relationship could be seen more clearly.

Using $P^{\prime}_d(k^{\prime})dk^{\prime} \sim P_m(M_B)dM_B$ and
$k^{\prime}\sim M_B^{\theta}$, we obtain the degree exponent of the
renormalized network to be $\gamma^{\prime} = 1+(\eta-1)/\theta.$ In
short, we argue that Eq.~(\ref{gammaprimesimple}) should be generalized to
\be
\gamma^{\prime}=
\begin{cases}
\gamma, & \text{for}~ \gamma \le \eta,\\
1+{{(\eta-1)}/\theta}, & \text{for}~\gamma > \eta,
\end{cases}
\label{gammaprime} \ee where $\theta \ne 1$. Accordingly, the
self-similarity holds even for $\gamma > \eta$ when \be
\theta=(\eta-1)/(\gamma-1).
\label{ss_condition} \ee  For $\ell_B=2$, we found that $\eta
\approx \gamma$ and $\theta \approx 1$; therefore,
$\gamma^{\prime}\approx \gamma$. For $\ell_B=5$, even though $\theta
\ne 1$, plugging $\theta \approx 0.5 \sim 0.6$ and $\eta \approx
1.8$ into Eq.~(\ref{gammaprime}), we obtain
$\gamma^{\prime}\approx 2.3 \sim 2.6$, which is in reasonable
agreement with $\gamma\approx 2.3$. Thus, self-similarity also
holds [Fig.~\ref{www}(b)]. We recall the previous study for the
fractal model with $\gamma=2.3$~\cite{newjournal}, finding that the
numerical results are the same as those of the WWW, as shown in
Figs.~\ref{www}(c) and (d).

\begin{figure*}
\centerline{\epsfxsize=12cm \epsfbox{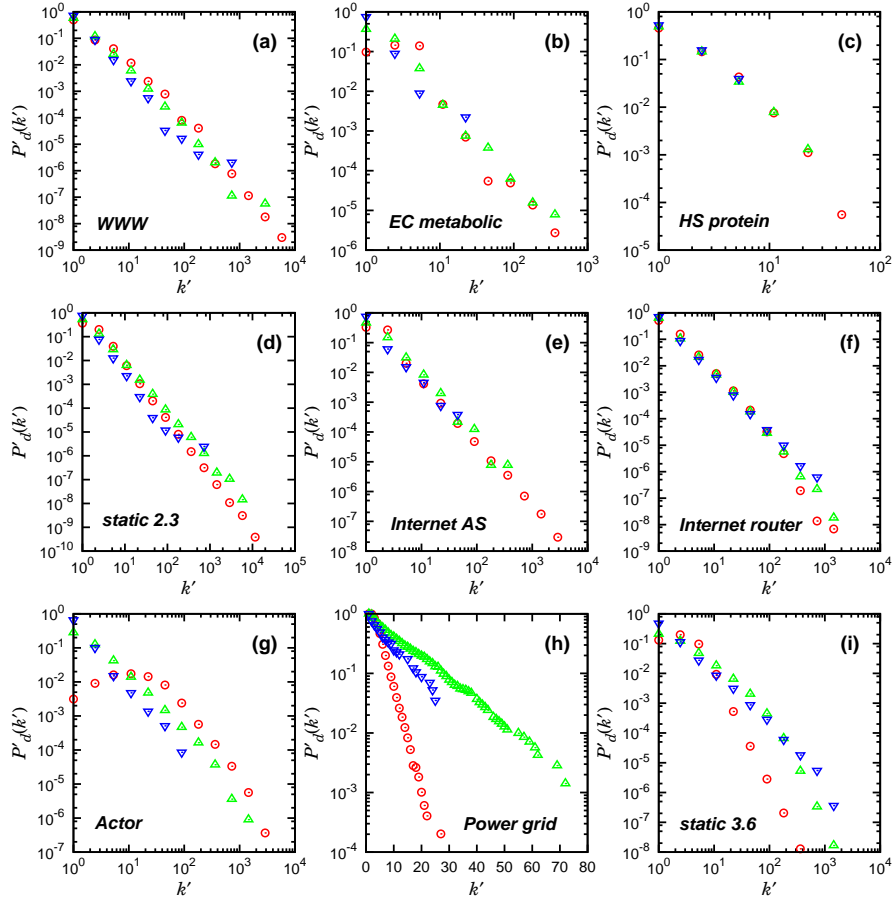}} 
\caption{(Color online) Renormalized degree distributions of real-world, (a)--(c) and
(e)--(h), and model, (d) and (i), networks under successive CG
transformations with a fixed box size $\ell_B=2$. Symbols represent
the original network (\textcolor{red}{$\circ$}), and the renormalized
networks after the first (\textcolor{green}{$\triangle$}) and the
second (\textcolor{blue}{$\triangledown$}) iterations. The networks
of (a)--(c) are fractals, and those of (d)--(i) are non-fractals.
The networks of (a)--(f) are self-similar while those of (g)--(i)
are not.}
\label{s4b}
\end{figure*}

The self-similarity condition in Eq.~(\ref{ss_condition}) is also
fulfilled for some non-fractal networks. The Internet at the
autonomous system (AS) level is a prototypical non-fractal SF
network. However, it exhibits self-similarity. We obtain 
$\eta\approx 1.8$ and the nonlinear relationship in Eq.~(\ref{nonlinear})
with $\theta \approx 0.7$, so that $\gamma\approx
\gamma^{\prime}\approx 2.1(1)$ for $\ell_B=2$, as shown in
Fig.~\ref{internet}, satisfying the condition in Eq.~(\ref{ss_condition}).

A few deterministic SF networks have been introduced, which were
constructed recursively starting from each basic structure. The
pseudofractal model~\cite{mendes} introduced by Dorogovtsev {\it et
al.}, the geometric fractal model introduced by Jung {\it et
al.}~\cite{jung}, and the hierarchical model~\cite{hmodel} are such
examples. These models restore their shapes under the CG achieved by
decimating the vertices with the smallest degree at each stage.
Thus, they are self-similar in shape. In those methods, the degree
distribution is scale invariant under the CG. However, they are not
fractal because fractal scaling, Eq.(\ref{fractal}), is absent and
the fractal dimension cannot be defined. The three deterministic
models~\cite{mendes,jung,hmodel} satisfy the self-similarity
condition in a trivial manner: $\eta=\gamma$ and $\theta=1$ under
the decimation. Here, the group-mass distribution is measured as the
distribution of the number of deleted vertices connected to each
coarse-grained vertex at each decimation stage. Thus,
$\gamma^{\prime}=\gamma$, and the models are self-similar, even
though they do not follow the fractal scaling, Eq.~(\ref{fractal}).

We also examine the change of the degree distributions under
successive CG transformations for the fractal and some non-fractal
networks in Fig.~\ref{s4b}. We confirm that the self-similarity
holds even for the non-fractal networks (d)--(f). It is noteworthy
that the relation in Eq.~(\ref{nonlinear}) is linear for the first
renormalization, but it becomes nonlinear for the second
renormalization as $\theta \approx 0.75$ and $\eta\approx 1.8$ for
the WWW. It is also interesting to note that the static model with
$\gamma=2.3$ is self-similar in Fig.~\ref{s4b}(d) while 
that with $\gamma=3.6$ in Fig.~\ref{s4b}(i) is not.
That is due to the topological difference of the SF networks with
$\gamma > 3$ and $2 < \gamma <3$ \cite{jskim,skeleton}.
When $\gamma$ is small, the edges are compactly concentrated around
the hubs while as $\gamma$ grows the edges more globally interweave
the network. Consequently, when renormalization is performed on the
SF network with $2 < \gamma <3$, only the nodes around each hub in
the original network are grouped into a supernode in a coarse-grained
network, and the supernode again becomes a hub with a corresponding size.
On the other hand, when $\gamma >3$, the nodes far from the hubs in
the original network have more chances to be connected to hubs via
global edges, and the supernodes in the coarse-grained network
become far bigger hubs than those in the original network.
The result is more heavily-tailed degree distributions in the
coarse-grained networks, as seen in Fig.~\ref{s4b}(i).

\section{CG by community structure}

Most SF networks in the real world contain functional modules or
community structures within them, which are organized in a
hierarchical manner. While the group-mass distributions are known to
follow a power law in Eq.~(\ref{gsize}), for many cases, however, they
exhibit a crossover between two distinct power-law behaviors or the
power-law behavior appears only in a limited range of mass. Also, they
are sensitive to various clustering algorithms~\cite{albert}. Thus,
it is not easy to find sufficiently good examples of clustered
networks with appropriate clustering algorithms to test our argument.

\begin{figure}[t]
\centerline{\epsfxsize=6.5cm \epsfbox{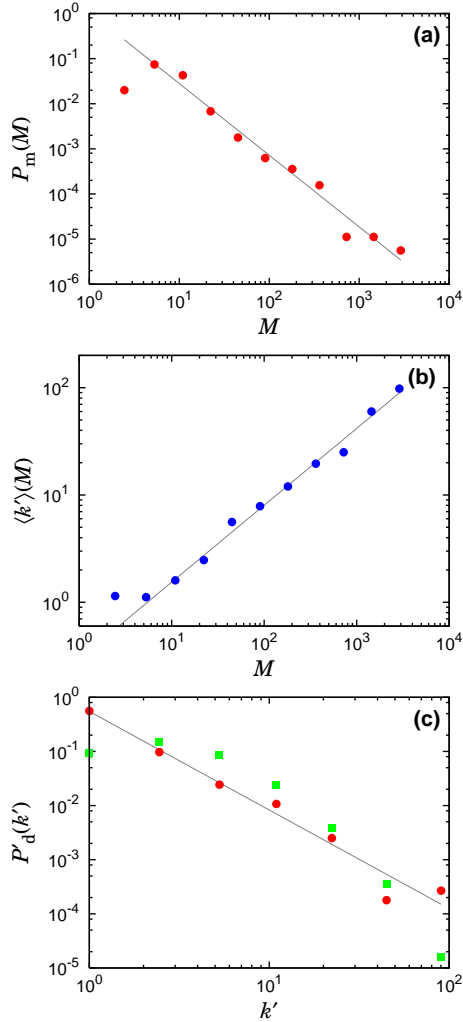}} 
\caption{(Color online) When the coauthorship network is clustered, (a) the box-mass
distribution, (b) the average renormalized degree $\langle k^{\prime}
\rangle$ versus group mass $M$, and (c) the degree distribution of the
original (\textcolor{green}{$\blacksquare$}) and renormalized
(\textcolor{red}{$\bullet$}) networks are plotted. Groups are
obtained by using the clustering algorithm of Ref.~\cite{clauset}. The solid
lines, drawn for reference, have slopes of (a) $-1.6$, (b) 0.7, and (c) $-1.8$.}
\label{fig:condmat}
\end{figure}

Here, we choose the algorithm proposed by Clauset {\it et al.}~\cite{clauset} 
and apply it to the cond-mat coauthorship
network~\cite{condmat}. The network data contain 13,861 vertices and
44,619 edges. Unfortunately, the degree distribution of this network
is not a power law. Nonetheless, the data are clustered into 175
groups obtained at the point where the modularity becomes maximum
in the clustering algorithm. The group-mass distribution is likely
to follow the power law in Eq.~(\ref{gsize}), and the exponent is estimated
to be $\eta\approx 1.6 \pm 0.2$ in Fig.~\ref{fig:condmat}(a). Next,
the CG is carried out; then, the degree distribution of the
renormalized network is examined. It shows a power-law behavior with
exponent $\gamma^{\prime}\approx 1.8 \pm 0.2.$ To check the formula of
Eq.~(\ref{gammaprime}), we measure the relationship of Eq.~(\ref{nonlinear})
between the renormalized degree and the group mass in
Fig.~\ref{fig:condmat}(b). The exponent $\theta$ is measured to be
$\theta\approx 0.7 \pm 0.1$. We plug the numerical values of $\eta$
and $\theta$ into the formula, and obtain $\gamma^{\prime}\approx
1.9\pm 0.4$, which is in reasonable agreement with the measured
value $\gamma^{\prime}\approx 1.8 \pm 0.2$ in
Fig.~\ref{fig:condmat}(c). The degree distribution of the original
network is overall skewed, while that of the renormalized network
follows the power law. Thus, self-similarity does not hold for
this case.

\section{Discussion and conclusions}

For the fractal networks, Song {\em et al.}~\cite{ss} showed that in
the box-covering method, the renormalized degree $k^{\prime}$ scales
as $k^{\prime} \sim s(\ell_B)k_{m}$, where $s(\ell_B)\sim
\ell_B^{-d_k}$ with $d_k={d_B/(\gamma-1)}$ and $k_{m}$ being the
largest degree in a given box. This form is not directly applicable
to non-fractal networks, but a modified form of $k_{\rm
max}^{\prime}\sim (N^{\prime}/N)^{1/(\gamma-1)}k_{\rm max}$ can be
applied for non-fractal but self-similar SF networks, where
$k_{\rm max}^{\prime}$ ($k_{\rm max}$) is the largest degree in the
entire renormalized (original) network and $N^{\prime}(N)$ is the
total number of nodes after (before) CG. The above relation can be
easily derived by using the scaling of the natural cutoff of degree,
$k_{\rm max}\sim N^{1/(\gamma-1)}$ and $k_{\rm max}^{\prime}\sim
N^{\prime 1/(\gamma-1)}$. Thus, for SF networks, it is more general
to formulate a scaling function in terms of the ratio $N^{\prime}/N$
rather than the length scale $\ell_B$.

Although, in this paper, we limited the notion of self-similarity to the scale
invariance of the degree distribution, one may wonder
if other quantities, such as $C(k)$ and $\langle k_{\rm
nn}\rangle(k)$, are scale invariant under the CG. We find that such
quantities tend to obey the scale invariance for the WWW in the box-covering 
method, but the statistics from real-world networks are not
sufficiently good to support this conclusion. That means, the self-
similarity thus defined does not imply any recursive topological identity
nor does it even guarantee that the degree distribution within each
group is identical from group to group. On the other hand, the origin of
fractality is understood by the power-law relation between the length
scale and the size of the skeleton underlying the original network \cite{jskim}.
In such senses, it seems that the fractality describes an important
topological feature of SF networks at a fundamental level while the
self-similarity does not. Moreover, the results we obtained till now
empirically show that all the fractal networks are self-similar, but
the converse is not true. Thus, we conjecture that fractality implies self-similarity.

In summary, we have studied the renormalization-group transformation
of the degree distribution, in particular, when the numbers of
renormalized vertices with each block follows a power-law
distribution, $P_m(M)\sim M^{-\eta}$. We found that the average
renormalized degree scales with the box mass as $\langle
k^{\prime}\rangle (M_B)\sim M_B^{\theta}$. The two exponents $\eta$
and $\theta$ can be nontrivial as $\eta \ne \gamma$ and $\theta \ne
1$. They act as relevant parameters and are analogous to the scaling
exponents associated with the magnetization and the singular part of
the internal energy in the renormalization group theory. We obtained the
degree exponent $\gamma^{\prime}$ of a remormalized network in terms
of $\eta$ and $\theta$. Many non-fractal networks are self-similar.
The notions of fractality and self-similarity are disparate in
SF networks, which is counterintuitive in view of their equivalence
in Euclidean space.

\begin{acknowledgments}
This work was supported by a Korean Science and Engeneering Foundation (KOSEF)
grant funded by the Korean Ministry of Science and Tecnology (MOST) (No.~R17-2007-073-01000-0). 
J.S.K.\ is supported by the Seoul Science Foundation. K.-I.G.\ is supported by a Korea University grant.
\end{acknowledgments}


\begin{references}
\bibitem{kadanoff} L. P. Kadanoff, Physics {\bf 2}, 263 (1966).
\bibitem{nijs} M. den Nijs, Physica A {\bf 251}, 52 (1998).
\bibitem{feder} J. Feder, {Fractals} (Plenum, New York, 1988).
\bibitem{ba} A.-L. Barab\'asi and R. Albert, {Science} {\bf 286}, 509 (1999).
\bibitem{hmodel} E. Ravasz and A.-L. Barab\'asi, Phys. Rev. E {\bf 67}, 026112 (2003).
\bibitem{newman} M. E. J. Newman, Phys. Rev. E {\bf 69}, 066133
(2004).
\bibitem{albert} L. Danon, A. D\'iaz-Guilera, J. Duch, and A.
Arenas, J. Stat. Mech. P09008 (2005).
\bibitem{ss} C. Song, S. Havlin, and H. A. Makse, {Nature (London)} {\bf 433}, 392
(2005).
\bibitem{jskim} J. S. Kim, K.-I. Goh, G. Salvi, E. Oh, B. Kahng, and D. Kim, Phys. Rev. E {\bf 75}, 016110 (2007).
\bibitem{chaos} J. S. Kim, K.-I. Goh, B. Kahng, and D. Kim, Chaos {\bf
17}, 026116 (2007).
\bibitem{newjournal} J. S. Kim, K.-I. Goh, B. Kahng, and D. Kim, New
J. Phys. {\bf 9}, 177 (2007).
\bibitem{bjkim} B. J. Kim, Phys. Rev. Lett. {\bf 93}, 168701
(2004).
\bibitem{bjkim_jkps} B. J. Kim, J. Korean Phys. Soc. {\bf 46}, 722
(2005).
\bibitem{rosenfeld} A. F. Rozenfeld, R. Cohen, D. ben-Avraham, and S. Havlin, Phys. Rev. Lett.
{\bf 89}, 218701 (2002).
\bibitem{mendes} S. N. Dorogovtsev, A. V. Goltsev, and J. F. F.
Mendes, Phys. Rev. E {\bf 65}, 066122 (2002).
\bibitem{jung} S. Jung, S. Kim, and B. Kahng, Phys Rev. E {\bf 65}, 056101 (2002).
\bibitem{static_model} K.-I. Goh, B. Kahng, and D. Kim,
Phys. Rev. Lett. {\bf 87}, 278701 (2001).
\bibitem{dslee} K.-I. Goh, D.-S. Lee, B. Kahng, and D. Kim, Phys. Rev. Lett. {\bf 91}, 148701 (2003).
\bibitem{skeleton} D.-H. Kim, J. D. Noh, and H. Jeong, {Phys. Rev. E} {\bf 70}, 046126 (2004).
\bibitem{goh2006} K.-I. Goh, G. Salvi, B. Kahng, and D. Kim, Phys. Rev. Lett. {\bf 96},
018701 (2006).
\bibitem{clauset} A. Clauset, M. E. J. Newman, and C. Moore, Phys.
Rev. E {\bf 70}, 066111 (2004).
\bibitem{condmat} M. E. J. Newman, Proc. Natl. Acad. Sci. USA {\bf 98,} 404 (2001).
\end{references}
\end{document}